\documentclass[aps,twocolumn,prd,showpacs,showkeys,preprintnumbers,superscriptaddress,nobibnotes,floatfix,longbibliography,nofootinbib]{revtex4-1}

\pdfoutput=1

\usepackage[english]{babel}
\usepackage{amsmath,amssymb,amsbsy,amstext,amsthm,amsfonts,array,booktabs,exscale,relsize,slashed,graphicx,upgreek,xcolor}
\usepackage{multirow,dcolumn,bm,enumerate,url,hyperref}
\usepackage[capitalise]{cleveref}
\usepackage{fontawesome}
\usepackage{mathrsfs}
\usepackage{longtable}
\usepackage{comment}
\usepackage{balance}

\definecolor{lightsabergreen}{rgb}{.14,.64,.14}
\definecolor{lightgreen}{rgb}{.14,.44,.14}

\hypersetup{
  colorlinks   = true, 
  urlcolor     = lightsabergreen, 
  linkcolor    = lightsabergreen, 
  citecolor   = lightsabergreen 
}

\begin{document}
\preprint{
\begin{minipage}{5cm}
\small
\flushright
UCI-HEP-TR-2024-06
\end{minipage}}

\title{Radiative corrections to light thermal pseudo-Dirac dark matter}

\author{Gopolang Mohlabeng}
\thanks{{\scriptsize Email}: \href{mailto:gmohlabe@sfu.ca}{gmohlabe@sfu.ca}}
\affiliation{Department of Physics and Astronomy, University of California, Irvine, California 92697, USA}
\affiliation{Department of Physics, Simon Fraser University, Burnaby, British Columbia, V5A 1S6, Canada}

\author{Adreja Mondol}
\thanks{{\scriptsize Email}: \href{mailto:amondol@uci.edu}{amondol@uci.edu}}
\affiliation{Department of Physics and Astronomy, University of California, Irvine, California 92697, USA}

\author{Tim M.P. Tait}
\thanks{{\scriptsize Email}: \href{mailto:ttait@uci.edu}{ttait@uci.edu}}
\affiliation{Department of Physics and Astronomy, University of California, Irvine, California 92697, USA}

\date{\today}

\begin{abstract}
Light thermal dark matter has emerged as an attractive theoretical possibility and a promising target for discovery at experiments in the near future.  Such scenarios generically invoke mediators with very small couplings to the Standard Model, but with moderately strong couplings within the dark sector, calling into question theoretical estimates based on the lowest order of perturbation theory.
As an example, we focus on a scenario in which (pseudo-)Dirac fermion dark matter is connected to the Standard Model via a dark photon charged under a new $U(1)^{\prime}$ extension of the standard model, and we investigate the impact of the next-to-leading order corrections to annihilation and scattering.
We find that radiative corrections can significantly impact model predictions for the relic density and scattering cross-section, depending on the strength of the dark sector coupling and ratio of the dark matter to mediator mass. We also show why factorization into the yield parameter $Y$ typically presented in literature leads to imprecision. 
Our results are necessary to accurately map experimental searches into the model parameter space and assess their ability to reach thermal production targets.
 
\end{abstract}

\maketitle
\section{Introduction \label{sec:intro}}
Of all of the viable scenarios of dark matter (DM), one of the most compelling is that of particles that are in thermal equilibrium with the Standard Model (SM) bath in the early Universe.
DM with a thermal history is very well motivated because it not only provides a feasible prediction for large non-gravitational interactions between dark and ordinary matter, but it is also highly predictive in nature, leading to clear targets for experimental searches.
Generically, after the DM has frozen out and its relic density set, it maintains the same interaction strength with the SM today, predicting that it is likely to be observable at a variety of experiments on Earth \cite{Bertone:2018krk}.
A canonical example is that of weakly interacting massive particles (WIMPs), a class of heavy ($\sim$ GeV-TeV scale) particles that interact with the SM through a roughly electroweak strength force \cite{Cushman:2013zza, Cooley:2022ufh}.\\

However, despite decades of search, heavy ($\gtrsim $ GeV) WIMPs are yet to be discovered and their parameter space has become tightly constrained \cite{Cooley:2022ufh}.
This lack of a clear signal has motivated searches for other visions of DM, including models with masses in the MeV to GeV regime interacting with the SM via new undiscovered forces.
While this class of DM is no longer a WIMP in the traditional sense, it can still be in equilibrium at early times and thus represents a thermal target.
It is a very attractive prospect as it may provide a road map to a wide and rich dark sector.
The possibilities of simplified DM models are vast. However, one of the most appealing scenarios is that of DM interacting with the SM through a new vector boson, often referred to as a dark photon, corresponding to a $U(1)^{\prime}$ extension of the SM \cite{Fayet:1980ad, Fayet:1980rr, Holdom:1985ag}.\\
 
Thermal sub-GeV DM, through a dark photon portal is very well motivated and has been the subject of much exploration in the literature. 
In this scenario, MeV to GeV mass DM can be produced and detected at a variety of current and near-future low energy accelerator experiments \cite{Izaguirre:2014bca, Izaguirre:2015yja, Izaguirre:2015pva, Izaguirre:2015zva, Feng:2017drg, Izaguirre:2017bqb, MiniBooNEDM:2018cxm, Marsicano:2018glj, Berlin:2018jbm, Mohlabeng:2019vrz, Tsai:2019buq, Duerr:2019dmv, Berlin:2020uwy, Batell:2021aja, Batell:2021ooj, CarrilloGonzalez:2021lxm, NA64:2021acr, Gori:2022vri, Krnjaic:2022ozp, Mongillo:2023hbs, Abdullahi:2023tyk, Brahma:2023psr}. 
Moreover, ambient DM in the Solar System can scatter with electrons and nucleons in small scale direct detection experiments with low energy recoil thresholds \cite{Emken:2019tni, Bloch:2020uzh, Essig:2022dfa}.
In addition, accelerator and underground detector probes are well complemented by astrophysical and cosmological constraints \cite{An:2017ojc, Chang:2018rso, DeRocco:2019jti, Fiorillo:2024upk, 
Fitzpatrick:2021cij, Giovanetti:2021izc}.
All together, these probes make for an exciting experimental program that shows promise for discovering light thermal DM in the near future.\\

As an illustrative model, we focus on pseudo-Dirac dark matter interacting with the SM via a dark photon mediator.  While we present results for the case in which the dark photon's interactions with the SM arise entirely via kinetic mixing with the hypercharge boson, our results are actually more general and apply to any theory in which a pseudo-Dirac dark matter particle interacts with a vector mediator whose interactions with the SM are much weaker than those with the dark matter itself.  Current experimental constraints and projections on this model that target the dark matter thermal relic abundance focus on a dark sector coupling $\alpha_{D}$ that is stronger than the QCD coupling at the electroweak scale.  As a result, the dark sector is strongly coupled and relatively large corrections from higher orders of perturbation theory are expected.  In this work, we compute the next-to-leading (NLO) order corrections on both the thermal annihilation and late time scattering processes with electrons.  We focus on the NLO corrections at ${\cal O}(\alpha_{D})$ and neglect the presumably negligible higher order corrections from the tiny kinetic mixing.\\

We find that the NLO corrections can be as large as ${\cal O}(10\%)$ for parameters typically discussed in the literature, and are thus necessary to take into account when precisely mapping experimental searches into the parameter space, and when comparing them with thermal production milestones  that serve as a prime target for a currently growing intensity frontier program.  It is worth pointing out that various theoretical constraints on this model resulting from its strong coupling have been considered in the literature, e.g. from the running of the dark sector coupling and the breakdown of perturbation theory \cite{Davoudiasl:2015hxa, Reilly:2023frg}. \\

The rest of this article is organized as follows: In Sec.~\ref{sec:MD} we describe the reference model, Sec.~\ref{sec:annihilation} is devoted to a full description of the thermal annihilation cross-section, including both leading and next-to-leading order (NLO) processes.
We compute the scattering cross-sections with electrons relevant for direct detection searches in Sec.~\ref{sec:scattering}. Finally we conclude in Sec.~\ref{sec:concl}. The appendices provide some technical details related to the counterterms in the on-shell renormalization scheme.

\section{Pseudo-Dirac Dark Matter \label{sec:MD}}

The basic module we consider consists of two Weyl fermions which play the role of dark matter and are  paired by a Dirac mass, $m_\chi$. These fermions are neutral under the SM gauge groups but have equal and opposite charge $\pm 1$ under a gauged dark U(1)$^\prime$, with corresponding gauge boson $A'_\mu$.  
The U(1)$^\prime$ symmetry is spontaneously broken by the vacuum expectation value of a dark scalar $\phi$, generating a mass for
$A'_\mu$ and [assuming the $\phi$ charge under U(1)$^\prime$ is chosen appropriately] Majorana masses for the two Weyl fermions.
We further assume that there is kinetic mixing between the dark photon and the SM hypercharge interaction, induced by unspecified UV physics.\\

In the mass basis, linear combinations of the original Weyl fermions appear as Majorana fermions, with their mixing determined by the Majorana masses and $m_\chi$.
We follow the standard assumptions in the literature that the Majorana masses are much smaller than $m_\chi$.  In this limit, the mass splitting between the two Majorana states goes to zero, and the pair can be approximately described as a single ``pseudo-Dirac" state $\chi$.  
Strictly speaking, this limit is experimentally ruled out by bounds from the cosmic microwave background \cite{Slatyer:2015jla,CarrilloGonzalez:2021lxm}.  However, relatively small splittings can ameliorate these bounds \cite{CarrilloGonzalez:2021lxm}, and the pseudo-Dirac limit is a reasonable approximation to viable models over much of the parameter space.\\

Putting these ingredients together, the resulting theory is described by the SM Lagrangian supplemented by terms describing the dark matter, mediator and dark Higgs:
\begin{eqnarray}
{\cal L}_{\text{DM}} & =  & i \overline{\chi} \gamma^\mu D_\mu \chi - m_\chi \overline{\chi} \chi -\frac{1}{4} X_{\mu \nu} X^{\mu \nu}  + \frac{m^2_{A'}}{2} A_\mu' {A'}^\mu \nonumber \\ & &
+ \frac{\varepsilon}{2} X^{\mu \nu} B_{\mu \nu} + |D_{\mu} \phi|^{2} - V(\phi)
\end{eqnarray}
where $\chi$ is a Dirac fermion packaging both of the original Weyl fermions in the limit of zero mass splitting (and thus including both the dark matter and its heavier partner when the mass splitting is taken into account).  $D_\mu \equiv \partial_\mu - i g_D q_{A'} A'_\mu$ is the covariant derivative for field of U(1)$^\prime$ charge $q_{A'}$, $X_{\mu \nu}$ and $B_{\mu \nu}$ are the field strengths for the U(1)$^\prime$ and SM hypercharge bosons, respectively, and $\varepsilon$ characterizes the strength of the kinetic mixing. We normalize $q_{A'}=1$ from here on for $\chi$.
After symmetry breaking, the scalar field acquires a vacuum expectation value (VEV) $v_{\phi}$ and can be parametrized in the unitary gauge as $\phi \rightarrow (v_{\phi} + H_{D})/\sqrt{2}$, where $H_{D}$ is the dark Higgs boson.\\ 

Diagonalizing the interactions and assuming $m_{A'} \ll M_Z$, the resulting theory contains a dark photon $A^\prime$ interacting with coupling strength $g_D$ with the DM, $\chi$ and strength $\varepsilon\, Q_f\, e$ with SM fields of electric charge $Q_f$. 
The dark photon mass is given by $m_{A^{\prime}} = q_{\phi}g_{D}v_{\phi}$, with $q_{\phi}$ the charge of the dark scalar. We will assume $q_{\phi} = 2$ such that the dark Higgs can have interactions with the DM Weyl fermions, allowing its VEV to contribute to their Majorana masses.
For simplicity, we assume that the mixing between $\phi$ and the SM Higgs is negligible.  The mixing parameter is important when considering the phenomenology
of $H_D$, but does not play an important role in dark matter annihilation at NLO.

\section{Thermal Freeze-out \label{sec:annihilation}}

For $m_\chi \leq m_{A'}$, annihilation of $\overline{\chi} \chi$ is predominantly into pairs of SM fermions,
\begin{equation*}
    \chi \, (p_a)+\overline{\chi} \, (p_b) \rightarrow f\, (p_1)\,+\overline{f}\, (p_2)
\end{equation*}
where $p_{a,b}$ label the incoming DM and $p_{1,2}$ the outgoing final state fermion momenta.
This cross section controls both the cosmological relic abundance via freeze-out and the prospects for indirect detection today.  Both processes take place for dark matter with typically non-relativistic velocity $v$, and can be approximated by the leading ($s$-wave) term in the expansion in $v^2$.  The cross section is related to the matrix element via,
\begin{equation}
\langle \sigma v \rangle = \frac{v}{64\pi ~ s ~ |\vec{p}_a|^2} ~ \int_{t_0}^{t_1} dt ~
\times \overline{| {\cal M} |^2},
\end{equation}
where $\overline{| {\cal M} |^2}$ is the matrix element, summed/averaged over final/initial polarization states, and $s\,\equiv(p_a+p_b)^2$ and $t\,\equiv(p_a-p_1)^2$ are the usual Mandelstam variables. 
In the non-relativistic ($v \rightarrow 0$) limit $s \simeq 4 m_\chi^2 + m_\chi^2 v^2$, $|\vec{p}_a| \simeq m_\chi v / 2$, and $t_{0,1} \simeq m_f^2 - m_\chi^2 \mp m_\chi^2 \sqrt{1 - m_f^2 / m_\chi^2}$.  

For dark matter masses below $\lesssim$~GeV, the annihilation rate into hadrons may be inferred by making use of the fact that $A^\prime$ couples to the same electromagnetic current as the photon.  Thus, the effective cross section for annihilation into hadrons at a given center-of-mass energy can be related to the ratio $R$ measured in $e^+ e^-$ reactions \cite{ParticleDataGroup:2014cgo}:
\begin{equation}
    R(s)= \frac{\sigma \left( e^{+} e^{-}\rightarrow \text{hadrons} \right)}
    {\sigma \left( e^{+} e^{-}\rightarrow \mu^{+} \mu^{-} \right)} ,
    \label{eqn:R}
\end{equation}
at a given center-of-mass energy $\sqrt{s}$.  The cross section for $\chi \bar{\chi}$ to annihilate into hadrons is thus:
\begin{equation}
\sigma \left( \chi \bar{\chi} \rightarrow \text{hadrons} \right) = \sigma \left( \chi \bar{\chi} \rightarrow \mu^+ \mu^- \right) \times ~R \left( 4 m^2_\chi \right). \nonumber
\end{equation}
\subsection{Tree-level annihilation}

\begin{figure}[t]
    \vspace{5mm}
  \centering
\includegraphics[width=0.35\textwidth]{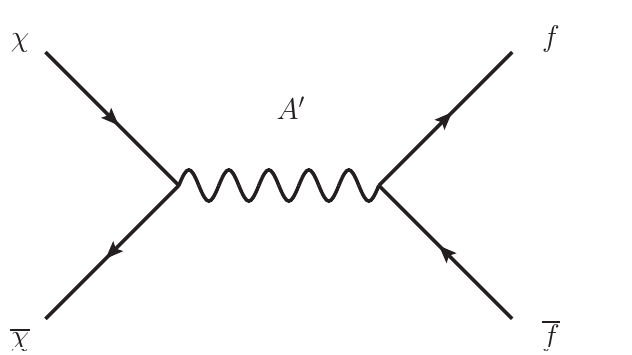}
  \caption{Tree-level Feynman diagram for DM annihilation into SM fermions}\label{fig:chi-f}
\end{figure}

At tree level, the annihilation into SM fermions proceeds via $s$-channel exchange of the $A^\prime$ (see Fig.~\ref{fig:chi-f}).  The leading order matrix element $\overline{| {\cal M} |^2}_{\text{LO}}$ is given by
\begin{eqnarray}\label{eqn:msq-tree}
& & 16 \pi^2 \alpha\, \alpha_D \,\varepsilon^2 Q_f^2 \times \\ & & 
\frac{(D-2) s^2 + 4 \left(m_f^2+m_\chi^2\right)^2 + \left( 4 s - 8 m_f^2 -8 m_\chi^2 \right) t + 4 t^2}
    {\left( s - m_{A'}^2 \right)^2 + \Gamma_{A'}^2 m_{A'}^2} \nonumber 
\end{eqnarray}
where $D \equiv 4 - \epsilon$ is the dimension of space-time, $\alpha_D=g_D^2/4\pi, \alpha=e^2/4\pi$ and  $\Gamma_{A'}$ is the width of the dark photon, given by
\begin{eqnarray}
\Gamma_{A'} &=&  
\Gamma (A' \rightarrow \chi \bar{\chi}) ~ \Theta(m_{A'}- 2 m_\chi) \nonumber \\ & & 
+~ \Gamma (A' \rightarrow e^+ e^-) ~ \Theta(m_{A'}- 2 m_e) \nonumber \\ & &
+~ \Gamma (A' \rightarrow \mu^+ \mu^-) \Theta(m_{A'}- 2 m_\mu) \nonumber \\ & &
+~ \Gamma (A' \rightarrow \mu^+ \mu^-) R( m^2_{A'}) ~ \Theta(m_{A'}- 2 m_\pi).
\label{eqn:decay-A-full}
\end{eqnarray}

Here,
\begin{eqnarray}
    \Gamma (A' \rightarrow \chi \bar{\chi}) = \frac{\alpha_D}{3} \,m_{A'} \left(1+ \frac{2 m_\chi^2}{m_{A'}^2}\right)\, \sqrt{1-\frac{4 m_\chi^2}{m_{A'}^2}}, \label{eqn:A-to-chi} \\
    \Gamma (A' \rightarrow f \bar{f}) = \frac{\alpha Q_f^2}{3}\, \varepsilon^2 \,m_{A'} \left(1+ \frac{2 m_f^2}{m_{A'}^2}\right)\, \sqrt{1-\frac{4 m_f^2}{m_{A'}^2}} ,
    \label{eqn:A-to-f}
\end{eqnarray}
are the partial decay widths into DM and SM fermions, respectively.
For much of the parameter space of interest, $m_{A'} \geq 2 m_\chi$ and $\varepsilon \ll 1$, leading to $\Gamma_{A'} \simeq \Gamma (A' \rightarrow \chi \bar{\chi})$.

In the non-relativistic limit, the thermally averaged tree-level cross section is 
\begin{equation} \label{eqn:sig-v-tree}
    \langle \sigma v \rangle_{\text{LO}} =
    \frac{8\pi\,\alpha\,\alpha_D\, Q_f^2 \, \varepsilon^2 
    \left( 2 m_\chi^2 + m_f^2 \right)}
    {\left(4 m_\chi^2 - m_{A'}^2\right)^2 + \Gamma_{A'}^2 m_{A'}^2}
    \,\sqrt{1-\frac{m_f^2}{m_\chi^2}}, 
\end{equation}
which agrees with Refs.~\cite{Izaguirre:2015zva,Kahn:2018cqs} in the $m_f \rightarrow 0$ limit.  

In the literature, it is common to factor out the combination $Y \equiv \varepsilon^2 \alpha_D (m_\chi / m_{A'} )^4$ which controls tree-level annihilation in the $m_{A'} \gg m_\chi$ limit.  This leads to imprecision.  First, the relevant parameter space is typically $m_{A'} \simeq m_\chi$, and errors of order $m_\chi^2 / m^2_{A'}$ are typically substantial.  
Second, the higher order corrections to the annihilation rate considered below are ${\cal O} (\alpha_D^2)$, and thus do not factorize in the same way.  And finally, various constraints and experimental prospects do not themselves factorize in the same way, which can make comparison using it as a parameter rather misleading.
For this reason, we focus on the direct model parameters $\left\{ \varepsilon, \alpha_D, m_\chi, m_{A'} \right\}$ in our analyses.
\subsection{NLO corrections}

\begin{figure}[th]
        \includegraphics[width=0.23\textwidth]{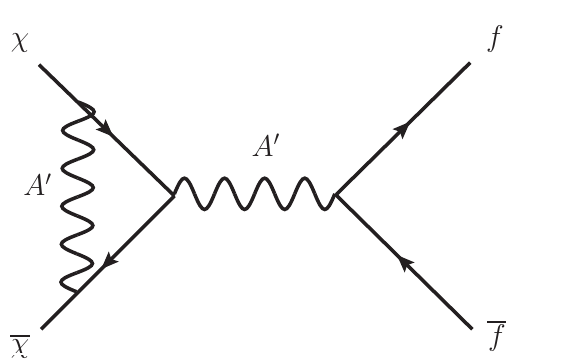}
         \includegraphics[width=0.23\textwidth]{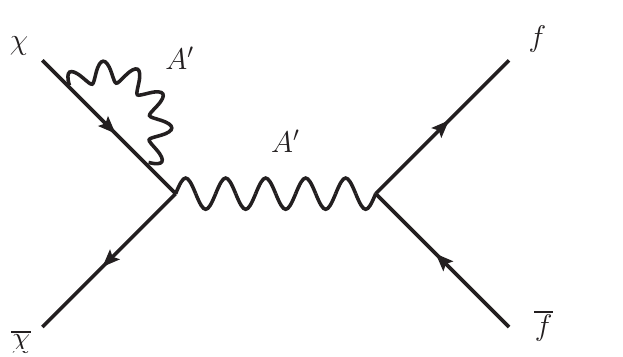}
         \includegraphics[width=0.23\textwidth]{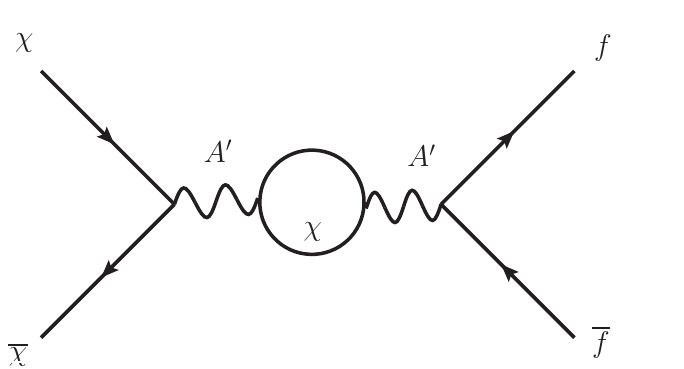}
     \caption{Representative Feynman diagrams for the one-loop corrections to $\chi \bar{\chi} \rightarrow f \bar{f}$.\label{fig:one-loop}}
\end{figure}

At ${\cal O} (\alpha_D^2)$, the annihilation cross section receives virtual corrections in the form of self-energy corrections to the dark photon propagator (these include diagrams with fermions and the dark Higgs in the loop) and incoming dark matter wave functions as well as a correction to the $A'$-$\chi$-$\overline{\chi}$ vertex (see Fig.~\ref{fig:one-loop}). In the regime of $m_A' \geq 2 m_\chi$, an additional emitted $A'$ would always be off-shell, leading the real emission corrections to be effectively higher order in $\varepsilon$ such that they can be safely neglected.  The ${\cal O} (\alpha_D^2)$ correction to the annihilation process is given by the interference between the leading order and next-to-leading order matrix elements in the non-relativistic limit, summed and averaged over the final and initial polarizations,
\begin{equation}
\overline{\delta \mathcal{M}^2} = 2\,\text{Re}\,\overline{\Bigg\{ \mathcal{M}^*_{\text{LO}} \times \mathcal{M}_{\text{NLO}} \Bigg\}} .
\label{eqn:msq-tot}
\end{equation}
We compute the one loop diagrams in the non-relativistic limit with the aid of \texttt{FeynCalc} \cite{Shtabovenko:2020gxv}, interfere them with the LO matrix elements, and reduce the resulting expressions via the Passarino-Veltman procedure \cite{Passarino:1978jh} to scalar integrals which we evaluate numerically using \texttt{LoopTools} \cite{Hahn:1998yk}. 
We cross-checked our results by hand as well as by using \texttt{Package-X} \cite{Patel:2016fam}. We find that the ultraviolet divergences cancel between the DM wave function corrections (see Appendix~\ref{sec:DMselfenergy}) and the vertex correction, as is expected based on the analogue of the Ward identity for U(1)$^\prime$.  
UV divergences in the one-loop correction to the dark photon propagator cancel against the $\delta Z_{A'}$ and $\delta m_{A^{\prime}}^2$ counterterms, renormalized in the on-shell scheme (see Appendix~\ref{sec:renormalization} for details), resulting in a final expression that is finite.

The resulting correction to the annihilation cross section from diagrams without a dark Higgs is expressed as 
\begin{widetext}
\begin{equation}
         \begin{aligned}
        \langle\delta\sigma v\rangle_{\text{NLO}}^{\text{DM}} &=-\frac{\alpha_D \, \langle \sigma v\rangle_{\text{LO}}}{3\,\pi}
            -\frac{\alpha_D \, \langle \sigma v\rangle_{\text{LO}}}{6\,\pi \, m_\chi^2 \left(4 \,m_\chi^2-m_{A'}^2\right)} 
            \Bigg\{
            -2 \left(m_{A'}^2-4 m_\chi^2\right) \\
            & \Bigg(m_\chi^2 \left(m_{A'}^2+2 m_\chi^2\right) \left(2\, B_0'\left(m_{A'}^2; m_\chi^2,m_\chi^2\right) +3 \,B_0'\left(m_\chi^2;m_{A'}^2,m_\chi^2\right)\right)
             +A_0\left(m_{A'}^2\right)-A_0\left(m_\chi^2\right)\Bigg)\\
            & +24 \,m_\chi^4 \,B_0\left(m_{A'}^2;m_\chi^2,m_\chi^2\right)+\left(m_{A'}^4-14 m_{A'}^2 m_\chi^2+40 m_\chi^4\right) B_0\left(m_\chi^2;m_{A'}^2,m_\chi^2\right)\\
            & +\left(m_{A'}^4+4 \,m_{A'}^2 m_\chi^2-56 \,m_\chi^4 \right) B_0\left(4 \,m_\chi^2;m_\chi^2,m_\chi^2\right)\\
            & +\left(m_{A'}^6+4\,m_{A'}^4 m_\chi^2-20\,m_{A'}^2 m_\chi^4-48\,m_\chi^6\right) C_0\left(m_\chi^2,m_\chi^2,4 m_\chi^2;m_\chi^2,m_{A'}^2,m_\chi^2\right)
            \Bigg\}
\end{aligned}
\end{equation}
\end{widetext}
where $A_0$, $B_0$, and $C_0$ are scalar integral functions, and $B^\prime_0$ is the derivative of $B_0$ with respect to its first argument. 
The contribution from the one-loop correction involving the dark Higgs is
\begin{widetext}
\begin{equation}\label{eqn:sig-v-dark-higs}
\begin{aligned}
         \langle\delta\sigma v\rangle_{\text{NLO}}^{\text{Higgs}}&=\frac{\langle \sigma v \rangle _{LO}\,\alpha_D\, q_\phi^2}{24\, \pi\,  m_{A'}^4 m_\chi^2 \left(m_{A'}^2-4 m_\chi^2\right)} \Bigg\{ B_0\left(m_{A'}^2;m_{A'}^2,m_H^2\right) \Bigg[-8 m_\chi^2 \Bigg(6 m_{A'}^6\\
         &-3 m_{A'}^4 m_H^2 +m_{A'}^2 \left(m_H^4+4 m_H^2 m_\chi^2\right)-2 m_H^4 m_\chi^2\Bigg)\Bigg]\\
         & + B_0\left(4 m_\chi^2;m_{A'}^2,m_H^2\right)\Bigg[m_{A'}^4 \left(m_{A'}^4-2 m_{A'}^2 \left(m_H^2-20 m_\chi^2\right)+\left(m_H^2-4 m_\chi^2\right)^2\right) \Bigg]  \\
         & + B_0^{\prime}\left(m_{A'}^2;m_{A'}^2,m_H^2\right) \Bigg[ 4 m_{A'}^2 m_\chi^2 \left(12 m_{A'}^4-4 m_{A'}^2 m_H^2+m_H^4\right) \left(m_{A'}^2-4 m_\chi^2\right)\Bigg]\\
         &+ (m_{A'}^2-m_H^2)  \left(m_{A'}^2-4 m_\chi^2\right)^2 \Bigg[A_0\left(m_H^2\right)-A_0\left(m_{A'}^2\right)\Bigg]\Bigg\}
    \end{aligned}
\end{equation}
\end{widetext}

\begin{figure}[t]
    \centering
    \includegraphics[width=0.49\textwidth]{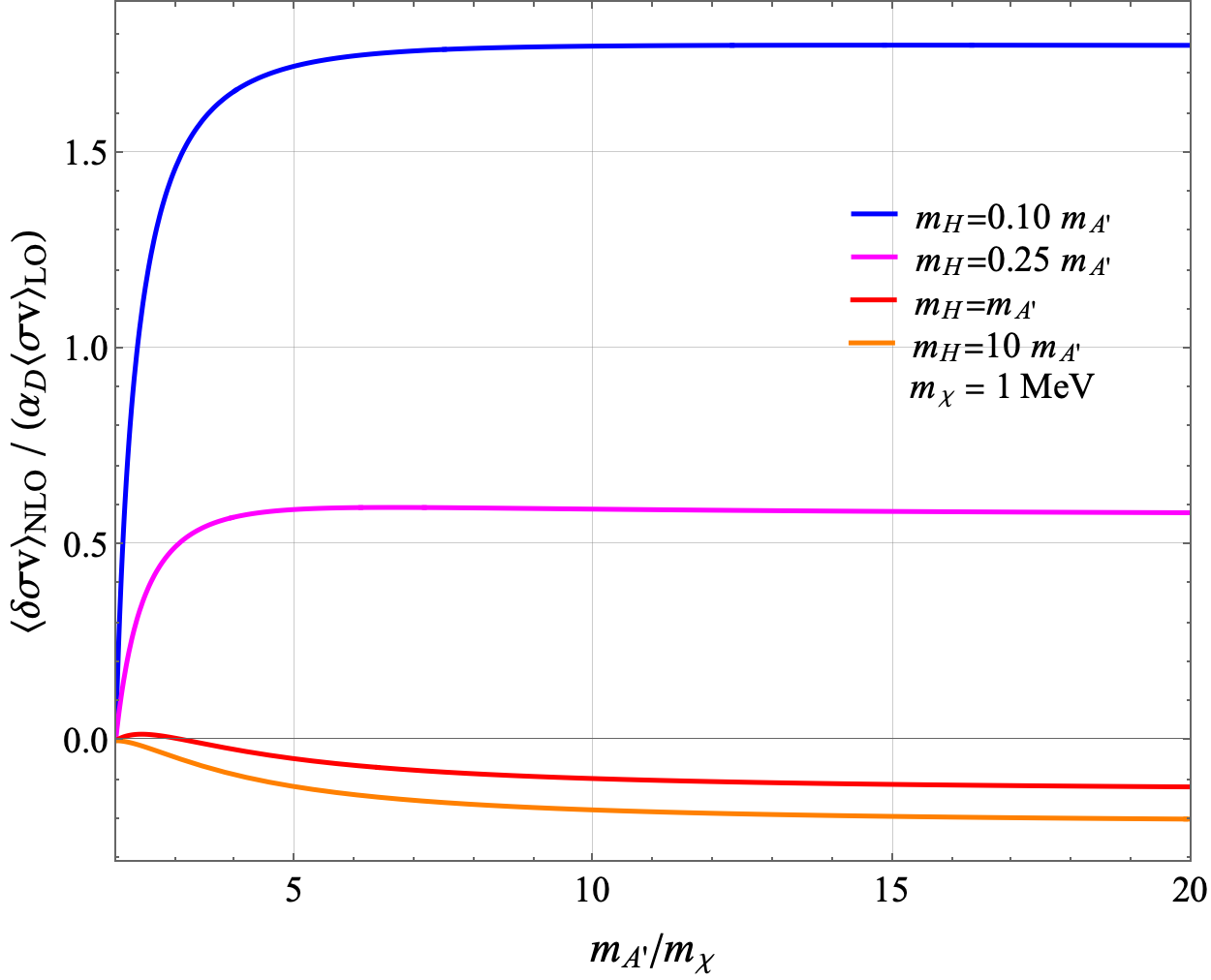}
    \caption{The ratio $\langle \delta \sigma v \rangle_{\text{NLO}} / (\alpha_D \langle \sigma v \rangle_{\text{LO}} )$ as a function of $m_{A'} / m_\chi$, for $m_\chi=1$ MeV and for different ratios of $m_H/m_{A'}=0.1,\, 0.25, 1 \text{ and } 10$, represented by the blue, magenta, red and orange lines, respectively.}  
    \label{fig:sigv-ratio}
\end{figure}

The full annihilation cross section to ${\cal O} (\alpha_D^2)$ is
\begin{equation}
    \langle \sigma v \rangle = \langle \sigma v \rangle_{\text{LO}} + \langle \delta \sigma v \rangle_{\text{NLO}}^\text{DM} + \langle\delta\sigma v\rangle_{\text{NLO}}^{\text{Higgs}} + {\cal O} \left( \alpha_D^3 \right).
    \label{eq:sigmav}
\end{equation}
%
%
The impact of the one-loop correction can be summarized by the quantity $\langle \delta \sigma v \rangle_{\text{NLO}} / (\alpha_D \langle \sigma v \rangle_{\text{LO}} )$, which characterizes the relative change compared to the leading order cross section with the $\alpha_D$ dependence scaled out. In Fig.~\ref{fig:sigv-ratio}, we plot this ratio as a function of the ratio of $m_{A'} / m_\chi$, for $m_\chi= 1$~MeV for different values of $m_H/m_{A'}$.
As illustrated in the figure, in the limit of large dark Higgs mass, the net effect of the NLO corrections is to reduce the cross section by a modest amount.
As $m_H/m_{A'}$ decreases, the NLO effect is to increase the total annihilation cross-section significantly. 
This trend saturates around $m_{A'}/m_\chi \sim 5$, after which the total cross-section flattens out for very large mass ratios. We also find that, provided $m_{A'}$ and $m_H$ are specified as a ratio to $m_\chi$, the quantity $\langle \delta \sigma v \rangle_{\text{NLO}} / (\alpha_D \langle \sigma v \rangle_{\text{LO}} )$ is insensitive to the mass of the dark matter over the entire range of interest, while we choose a benchmark $m_{\chi} = 1$ MeV, the results are identical for e.g., $m_{\chi} = 1$ GeV.

\subsection{Relic density}

To compute the relic density we follow the description of Refs.~\cite{Steigman:2012nb, Saikawa:2020swg} where it was pointed out that care is required in the treatment of the number of relativistic degrees of freedom as a function of temperature for sub-GeV dark matter annihilations.
Hence, we solve the Boltzmann equation for the comoving number density of dark matter,
\begin{equation}
\frac{dY}{dx} = \frac{s \langle \sigma v \rangle}{H x} \left[ 1 + \frac{1}{3} \frac{d ({\rm ln}~g_{s})}{d({\rm ln}~T)}\right] (Y_{eq}^{2} - Y^{2}).
\end{equation}
We refer the reader to Ref.~\cite{Steigman:2012nb} for discussion of the relevant quantities.
Ref.~\cite{Steigman:2012nb} presented results for the target cross section which reproduces the observed DM density only down to $m_{\chi} \sim$ 100 MeV. To cover the relevant parameter space, we implement their formalism, reproducing their results in the regime that they covered, and extend them down to $\sim$ 1 MeV DM masses.

\begin{figure*}[t!]
    \centering
    \includegraphics[width=0.49\textwidth]{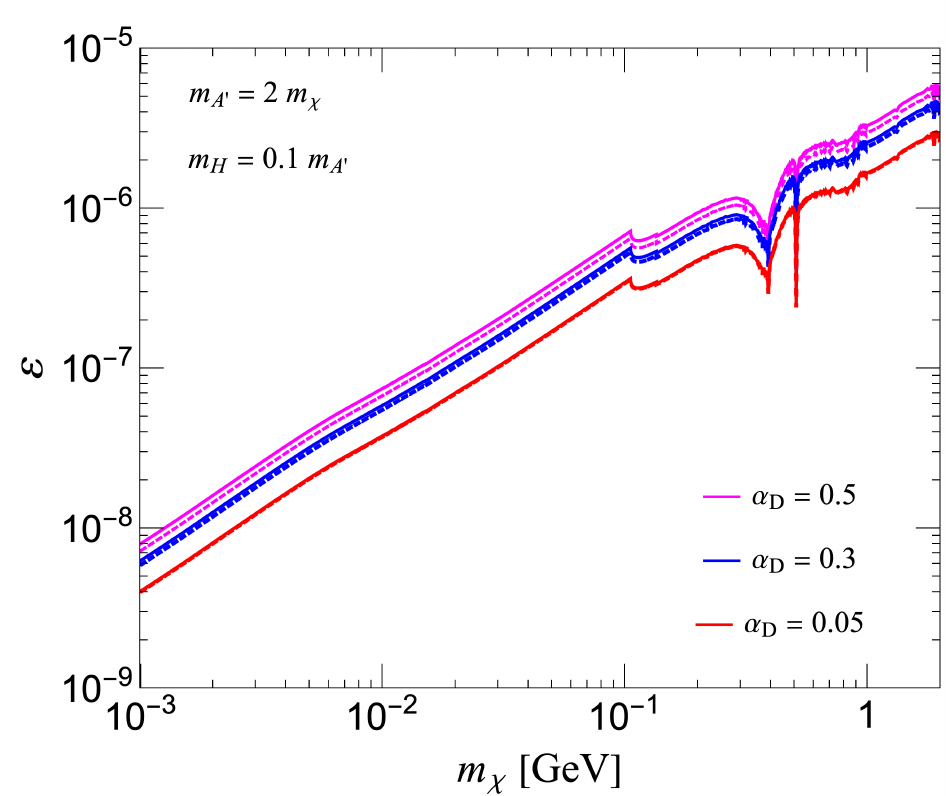}
    \includegraphics[width=0.49\textwidth]{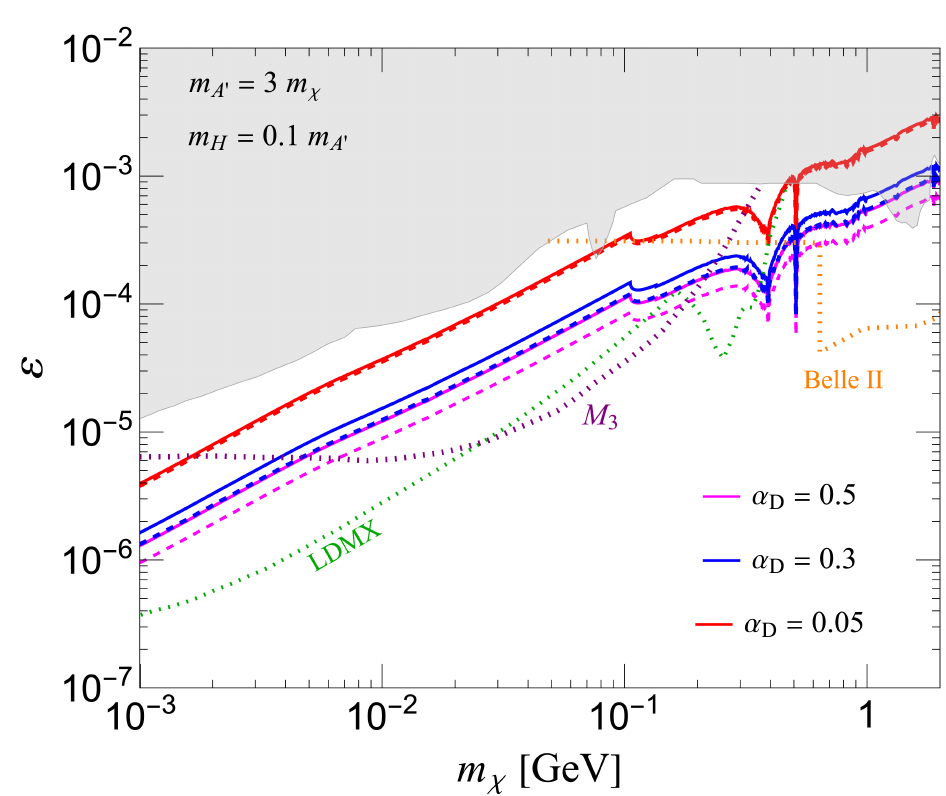}
    \includegraphics[width=0.49\textwidth]{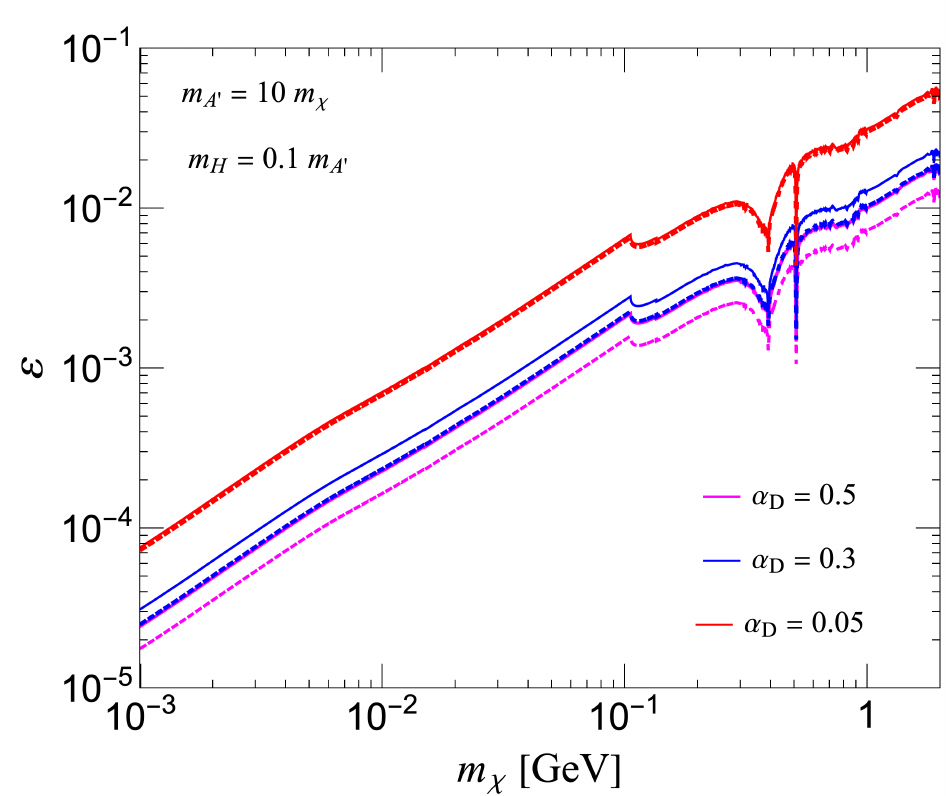}
    \caption{Value of $\varepsilon$ producing the observed relic abundance of dark matter as a function of $m_\chi$, for $\alpha_{D} = $ 0.05, 0.3 and 0.5 (red, blue and magenta lines respectively) based on LO only (solid lines) and NLO + LO (dashed lines) cross sections and for the indicated value of $m_{A'}/m_\chi$ and $m_H/m_{A^\prime}$ on each panel.  
    For the $m_{A^{\prime}} = 3 m_{\chi}$ panel, the gray shaded regions are the exclusions from past accelerator experiments, and the green, purple and orange dotted lines are projections from the upcoming LDMX, $M_{3}$ and Belle II experiments respectively~\cite{Gori:2022vri}. We show the experimental constraints in the $m_{A^{\prime}} = 3 m_{\chi}$ panel, which have been extensively studied in the literature. It is worth emphasizing that there are expected to be analogous constraints, on the other panels, but they require further analysis to be determined and will be addressed in future work.} 
    \label{fig:eps-mchi}
\end{figure*}

In Fig.~\ref{fig:eps-mchi} we show plots of kinetic mixing as a function of DM mass for a few representative values of the ratio of $m_{A^{\prime}}/m_{\chi}$ and choose a benchmark parameter point $m_{H} = 0.1 m_{A^{\prime}}$. These plots were obtained by scanning through the kinetic mixing for each DM mass and finding the combination ($\varepsilon$, $m_{\chi}$) resulting in the observed relic abundance $\Omega h^{2} = 0.12$ \cite{Aghanim:2018eyx}. Each plot shows the required kinetic mixing parameter for three different values of the dark sector coupling $\alpha_{D}$, corresponding to the LO only (solid lines) and NLO + LO (dashed lines) computations, respectively.
For $m_{A^{\prime}}/m_{\chi} = 3$, which is typically chosen as a benchmark parameter point in the literature and is the most extensively studied.
Hence, we show the current experimental constraints on this parameter space (a combination of both accelerator and astrophysical probes \cite{Gori:2022vri}) in the gray shaded region. 
In that panel, the green, purple and orange dotted lines are sensitivity projections for the upcoming LDMX, $M_{3}$ and Belle II experiments, respectively. 
Assessing the experimental constraints on the other two panels is complex, and a thorough analysis is beyond the scope of this work. 
Following the trend in Fig.~\ref{fig:sigv-ratio}, we see in the top right panel, i.e. $m_{A^{\prime}} = 2 m_{\chi}$, the loop corrections result in not-so-significant changes in the thermal relic density lines, especially when lowering the value of the dark sector coupling constant $\alpha_{D}$.
As one increases the value of $m_{A^{\prime}}/m_{\chi}$, we start seeing a significant change in the thermal relic line as a result of the loop corrections, especially for the large values of $\alpha_{D}$ which are often referenced in the literature, making these effects important to consider. We also note that the effect of these corrections can be even more dramatic if one further decreases the mass ratio $m_{H}/m_{A^{\prime}}$.
Finally, in order for the future experiments to correctly characterize their future signals, the significance of these results should be taken into account.
\section{Dark matter direct detection 
\label{sec:scattering}}

In this section we investigate NLO corrections to light pseudo-Dirac DM scattering with the SM in direct detection experiments. Due to crossing symmetry, the corrections we consider here are described by similar diagrams as in Fig.~\ref{fig:one-loop}. 
Since the dark photon kinetically mixes with the SM photon, DM can scatter universally with both leptons and nucleons.
Given the tiny momentum transfers and our parameter space of interest for which $m_{A^{\prime}} \gtrsim m_{\chi}$, the interaction can be approximated as an effective four-point fermion interaction, integrating out the mediator. 

The spin averaged matrix element squared in the zero-momentum exchange limit is
\begin{equation}
    |\overline{\mathcal{M}_{EFT}}|^{2} = \Bigg \{ \lvert \overline{\mathcal{M}} \rvert^2_{\text{LO}}+ \overline{\delta \mathcal{M}^2}\Bigg \}_{t \rightarrow 0}
\end{equation}
where $\overline{\delta \mathcal{M}^2}$ corresponds to Eq.~(\ref{eqn:msq-tot}) and includes contributions from the one-loop corrections to the vertex, DM self-energy and the dark photon vacuum polarization, analogous to those represented in Fig.~\ref{fig:one-loop}.

Given our focus on sub-GeV DM and the prospects for its detection at upcoming experiments, we restrict our attention to DM scattering with electrons.
However our general computations can be extended to nucleon scattering with the incorporation of the appropriate form factors for either spin-dependent or spin-independent scattering.\\

The differential DM-electron scattering cross-section as a function of the momentum transfer $q$ is usually written
\begin{eqnarray}
\frac{d \sigma}{d q^2} &=& \frac{\overline{\sigma}_{e}}{4 \mu v^{2}} |F_{DM}(q)|^{2}.
\end{eqnarray}
Here $\overline{\sigma}_{e}$ is defined as the free scattering cross-section at a reference value $q = \alpha m_{e}$ \cite{Essig:2011nj, Emken:2019tni}, which typifies the momentum of an electron bound in an atom in the detector.
For the DM mass range, and ratios with the mediator mass ($m_{A^{\prime}}/m_{\chi}$) we consider in this work, $q \sim \mu v \ll m_{A^{\prime}}$, and hence $F_{DM}(q) \rightarrow F_{DM}(q=0) \sim 1$. \\

Similarly to the annihilation case, the cross-section for scattering with electrons up to ${\cal O} (\alpha_D^2)$, can be parameterized as:
\begin{equation}
    \overline{\sigma_{e}} = \overline{\sigma}_{e}^{\rm LO} +  \overline{\delta\sigma_e}^{\rm NLO}_{\text{DM}} +\overline{\delta\sigma_e}^{\rm NLO}_{\text{Higgs}}+ {\cal O} \left( \alpha_D^3 \right). 
\end{equation}
At zero momentum transfer, the LO scattering cross-section is given by~\cite{Essig:2011nj, Emken:2019tni}
\begin{equation}\label{eqn:DirDet-sigmae-low}
    \overline{\sigma}_{e}^{\rm LO}\,=\,\frac{16\,\pi  \alpha\, \alpha_D\,\varepsilon ^2\, \mu ^2}{m_{A'}^4},
\end{equation}
where $\mu \equiv m_{\chi} m_{e}/(m_{\chi} + m_{e})$ is the reduced mass between the DM and the electron.
The ${\cal O} (\alpha_D^2)$ correction can be written in terms of scalar integrals as:
\begin{widetext}
    \begin{equation}
     \begin{aligned}
         \overline{\delta \sigma_e}^{\text{NLO}}_{\text{DM}}\,& = \frac{\alpha_D\,\overline{\sigma_e}^{LO}}{12\, \pi\,  m_{A'}^2\, m_\chi^2}\Bigg[-m_{A'}^2 \left(-2\, m_{A'}^2-4\, m_\chi^2\right) \Bigg\{4\, m_\chi^2 B_0^{\prime}\left(m_{A'}^2; m_\chi^2,m_\chi^2\right)\\
         &-3 \Bigg(B_0\left(0; m_\chi^2,m_\chi^2\right)-B_0\left(m_\chi^2; m_{A'}^2,m_\chi^2\right)\\
         &+\left(m_{A'}^2-2\, m_\chi^2\right) C_0\left(0,m_\chi^2,m_\chi^2;m_\chi^2,m_\chi^2,m_{A'}^2\right)-2\, m_\chi^2 B_0^{\prime}\left(m_\chi^2;m_{A'}^2,m_\chi^2\right)\Bigg)\Bigg\}\\
         &-16\, m_\chi^4 \Bigg( B_0\left(m_{A'}^2;m_\chi^2,m_\chi^2\right)-B_0\left(0;m_\chi^2,m_\chi^2\right)\Bigg)\Bigg]
\end{aligned}
\end{equation}
\end{widetext}

The contribution to the scattering cross-section from the dark Higgs is 
\begin{widetext}
    \begin{equation}
     \begin{aligned}
        \overline{\delta \sigma_e}^{\text{NLO}}_{\text{Higgs}}\,& = \frac{q_\phi^2\,\alpha_D\,\overline{\sigma_e}^{LO}}{6\, \pi\,  m_{A'}^4 (m_{A'}^2-m_H^2)} \Bigg[(m_{A'}^2-m_H^2) \Bigg\{m_{A'}^2 \Bigg(\left(m_{A'}^2-m_H^2\right)^2 B_0^{\prime}\left(0; m_{A'}^2,m_H^2\right)\\
         &+\left(12\, m_{A'}^4-4\, m_{A'}^2\, m_H^2+m_H^4\right) B_0^{\prime}\left(m_{A'}^2;m_{A'}^2,m_H^2\right)\Bigg)\\
         &-2 \left(6\, m_{A'}^4-3\, m_{A'}^2\, m_H^2+m_H^4\right) B_0\left(m_{A'}^2; m_{A'}^2,m_H^2\right)\Bigg\}\\
         &+2 \left(6\, m_{A'}^4-3\, m_{A'}^2 \,m_H^2+m_H^4\right) \Bigg(A_0\left(m_{A'}^2\right)- \text{A}_0\left(m_H^2\right)\Bigg) \Bigg]
\end{aligned}
\end{equation}
\end{widetext}
where we have used the identity $B_0 (0; m_1^2, m_2^2) = (A_0(m_1^2)-A_0(m_2^2))/(m_1^2-m_2^2)$.
\begin{figure*}[t!]
    \centering
    \includegraphics[width=0.48\textwidth]{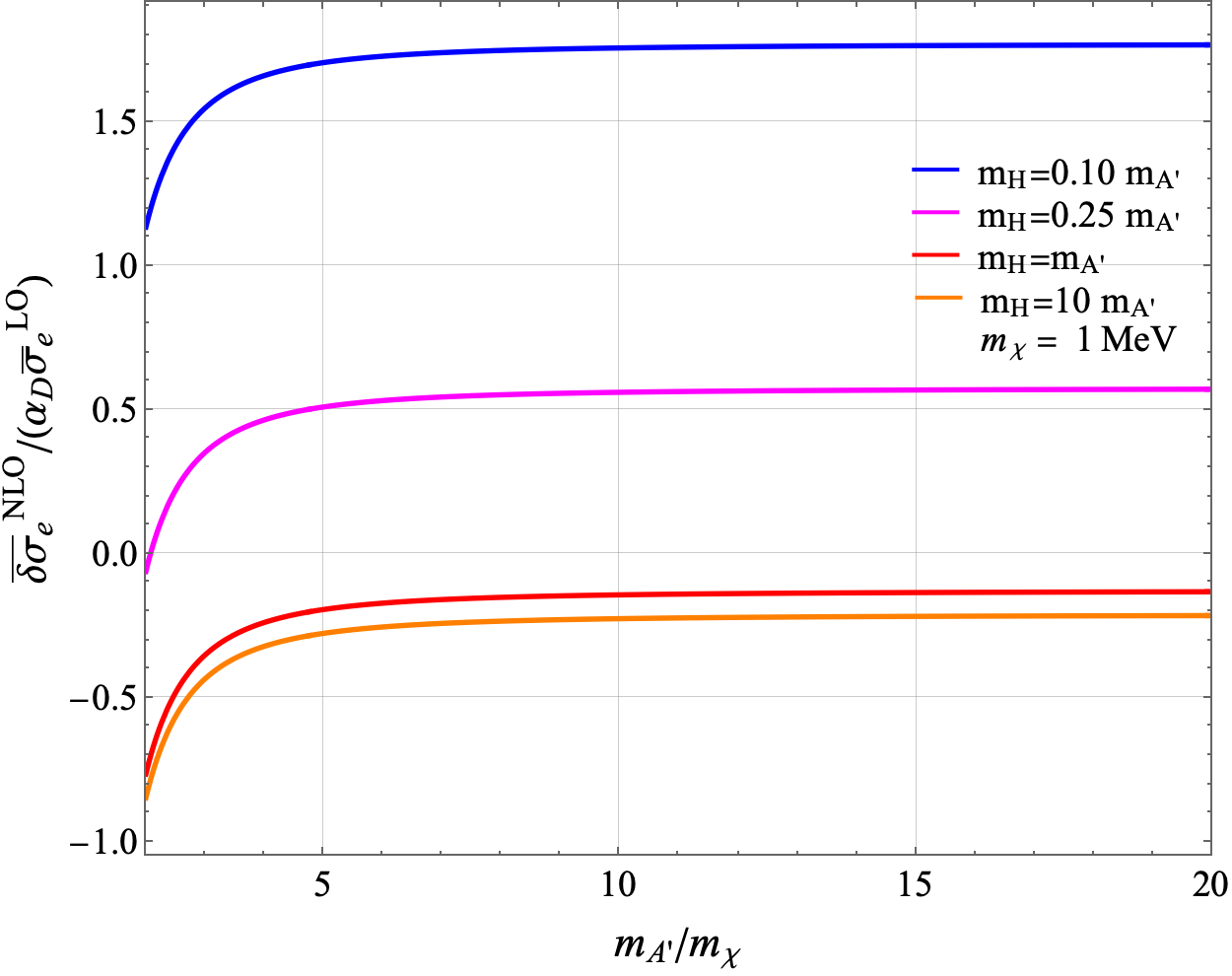}
    \includegraphics[width=0.48\textwidth]{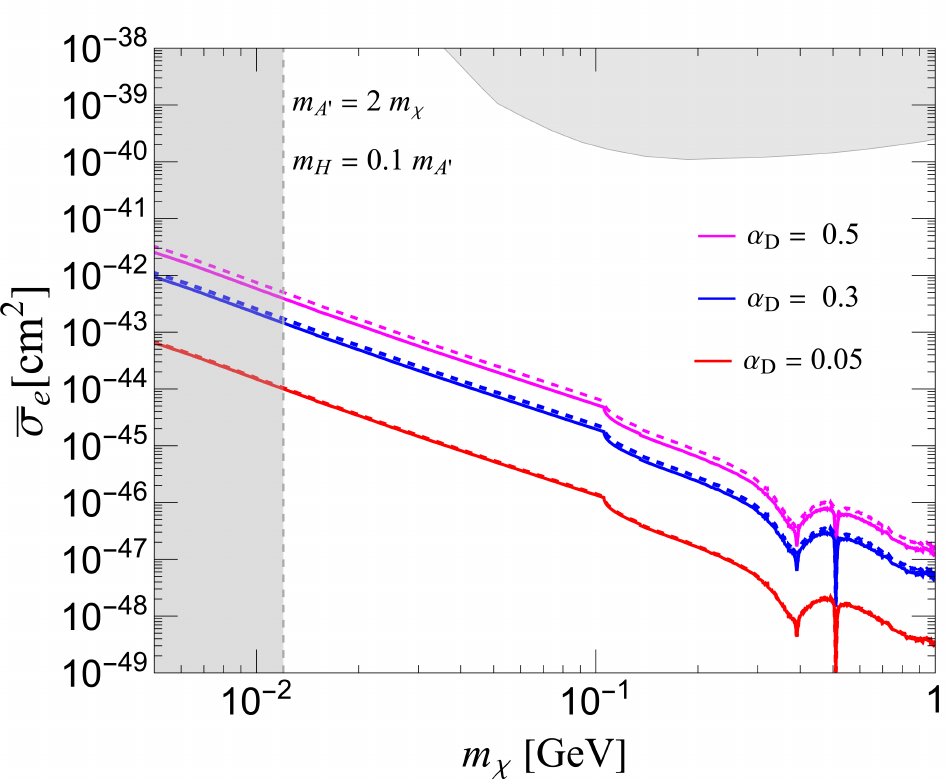}
    \caption{Dark matter direct detection parameter space. The left panel shows the ratio of NLO to LO electron scattering cross-section vs $m_{A^{\prime}}/m_{\chi}$ ratio for $m_{\chi} = 1$ MeV and for $m_H/m_{A^\prime} = 0.1, 0.25, 1 \text{ and } 10$, represented by the blue, magenta, red and orange lines, respectively.  The plot for $m_\chi = 1$ GeV is identical to the left panel. The right panel shows dark matter-electron scattering cross-section vs $m_{\chi}$ for dark matter thermal relic abundance with $\alpha_{D} = $, 0.05, 0.3 and 0.5, represented by the red, blue and magenta lines respectively, for $m_H/m_{A^\prime}=0.1$. We show both the LO only (solid lines) and NLO + LO (dashed lines) for the $m_{A^{\prime}}/m_{\chi} = 2$ case. The upper gray shaded region represents the current model independent direct detection constraints from a combination of the XENON1T, PANDAX and SENSEI experiments and the vertical gray shaded region is the constraint from $\Delta N_{\text{eff}}$. All constraints were obtained from Ref.~\cite{Essig:2022dfa}.} 
    \label{fig:sigmae-mchi}
\end{figure*}

In Fig.~\ref{fig:sigmae-mchi}, we show the direct detection constraints on the model. 
The left panel characterizes the importance of the one-loop correction, parametrized by the quantity $\overline{\delta\sigma}_{e NLO}/(\alpha_{D} \overline{\sigma}_{e LO})$ in terms of the ratio $m_{A^{\prime}}/m_{\chi}$ and for different values of $m_{H}/m_{A^{\prime}}$, similar to Fig.~\ref{fig:sigv-ratio}. 
Again, we find that this quantity is independent of the DM mass, and we choose $m_{\chi} = 1$ MeV for reference.

In the right panel, we show the averaged DM-electron scattering cross-section vs DM mass. The red, blue and magenta lines represent DM produced through the thermal freeze-out mechanism for benchmark choices of $\alpha_{D} = $ 0.05, 0.3 and 0.5 respectively. The dashed lines are NLO + LO contributions, while the solid lines are only at LO.
For illustration, we only show the case for $m_{A^{\prime}}/m_{\chi} \simeq $ 2 and $m_{H} = 0.1 m_{A^{\prime}}$. We see a clear dependence on the dark sector coupling $\alpha_{D}$ and differences between the NLO + LO and LO only contributions.

Also shown as the gray shaded region are the model independent constraints on this parameter space. 
Below $\sim$ 10 MeV, there is a stringent bound from big bang nucleosynthesis and $\Delta N_{\text{eff}}$, whereas the upper gray shaded region indicates bounds from a combination of direct detection experiments, including XENON1T, PANDAX and SENSEI (obtained from Ref.~\cite{Essig:2022dfa}).

Similarly to the annihilation case above, we see the significance of the radiative corrections for different values of the ratio $m_{H}/m_{A^{\prime}}$, particularly for large  $m_{A^{\prime}}/m_{\chi}$. For $m_{H}/m_{A^{\prime}} \geq 1$ the effect of a heavy dark Higgs is decreased.
As one decreases $m_{H}/m_{A^{\prime}}$, we notice a significant correction to the total scattering cross-section that increases as $m_{A^{\prime}}/m_{\chi}$ rises, saturating for $m_{A^{\prime}}/m_{\chi} \sim 5$.

\section{Discussion and Conclusions}
\label{sec:concl}
In this work, we revisit the computations for sub-GeV thermal dark matter processes. We consider a highly motivated and sought after scenario, in which (pseudo-)Dirac dark matter couples to the SM via a massive dark photon, representing a compelling benchmark model of dark matter. We focus on the
corrections of order $\alpha_D$, which is typically invoked to be large in order to obtain the observed relic density of the dark matter.\\

We compute the NLO corrections for thermal (pseudo-)Dirac dark matter annihilating in the early Universe, as this represents a very important target for upcoming experiments.
We find that the NLO corrections can be quite significant, much larger than 10s of percents, depending on the $m_{A^{\prime}}/m_{\chi}$ ratio, the mass of the dark Higgs and the strength of the dark sector coupling constant $\alpha_{D}$.
We find that this can result in a dramatic change compared to the 
targets inferred from tree-level calculations, as the mass of the Higgs in this model can be smaller compared to the mass of the dark photon. Hence it is an important result to establish the ability of proposed experiments to discover or constrain MeV scale dark matter, and to precisely quantify what we will learn from such experiments in the future.
Furthermore, our results show that factorization into the ``yield parameter'' $Y$, as is common practice in the literature, can be misleading, especially in the resonance region, where the NLO corrections are more significant.\\

For completeness, we also show the direct detection limits based on DM-electron scattering. 
We find that the NLO corrections are as important as for the case of annihilation.
One may wonder about the size of these corrections if the dark photon mass was generated through a different mechanism such as a Stueckelberg mass, and there were no dark Higgs particles.
The scenarios for $m_{H}/m_{A^{\prime}} \geq 1$ in the plots above correspond to this case. As we see, the corrections are still noticeable, though not as significant as those obtained for lighter Higgs masses.
We leave a more dedicated study of pseudo-Dirac dark matter with larger mass splittings as well as other Lorentz structures of interest to future studies.

\section*{Acknowledgements}
We thank Basudeb Dasgupta for useful discussions. 
G.M. acknowledges support from the UC office of the President through the UCI Chancellor's Advanced Postdoctoral Fellowship. G.M. is also supported in part by the National Sciences and Engineering Research Council of Canada.
T.M.P.T. and G.M. are supported in part by the U.S. National Science Foundation under Grant No. PHY-2210283.
This work was performed in part at Aspen Center for Physics, which is supported by National Science Foundation Grant No. PHY-2210452.

\appendix
\balance
\section{Dark Matter Self-Energy}
\label{sec:DMselfenergy}

In terms of the Passarino-Veltmann scalar functions, the dark matter self-energy takes the form
\begin{equation}
\begin{aligned}
    \Sigma (p)\, &= \frac{\alpha_D}{8 \pi} \Bigg[\frac{(\epsilon -2) \slashed{p}} {p^2}\Bigg(A_0\left(m_{A'}^2\right)- A_0\left(m_\chi^2\right)\\
    & + \left(-m_{A'}^2+m_\chi^2+p^2\right) B_0\left(p^2; m_{A'}^2,m_\chi^2\right)\Bigg)\\
     & -2 m_\chi (\epsilon -4) B_0\left(p^2; m_{A'}^2,m_\chi^2\right)\Bigg]~,
\end{aligned}
\end{equation}
where $\slashed{p} \equiv \gamma^\mu p_\mu$ and $\epsilon=\,4-D$.  The dark matter wave-function counter-term in the on-shell renormalization scheme, $\delta Z_2 = d\Sigma / d \slashed{p}$,$(\slashed{p} \rightarrow m_\chi)$, is given by:
\begin{equation}
\begin{aligned}
    \delta Z_2 &=
    -\frac{\alpha_D}{8 \pi \, m_\chi^2} \Bigg[(\epsilon -2) \Bigg(m_{A'}^2 \left(-B_0\left(m_\chi^2; m_{A'}^2,m_\chi^2\right)\right)\\
    & +A_0\left(m_{A'}^2\right)-A_0\left(m_\chi^2\right)\Bigg)\\
     & +2\, m_\chi^2 \left(m_{A'}^2 (\epsilon -2)-4 \,m_\chi^2\right) B_0'\left(m_\chi^2; m_{A'}^2,m_\chi^2\right)\Bigg]~.
\end{aligned}
\end{equation}

The dark matter mass counter-term in the on-shell scheme, $\delta m_\chi = \Sigma(p)$, $(\slashed{p} \rightarrow m_\chi)$ can be similarly extracted, but does not enter into the computation of annihilation or scattering with electrons at NLO.

\section{Dark Photon Self-Energy}
\label{sec:renormalization}

The tree level dark photon propagator in the unitary gauge takes the form:
\begin{equation}
     i\,D^{(0)}_{\mu\nu}(k^2)=\frac{i\, (-g_{\mu\nu}+k_\mu k_\nu / m_{A'}^2)}{k^2-m_{A'}^2}
     \label{eqn:prop-mA}
\end{equation}
where $k$ is the momentum of the $A'$. The self energy correction can be expressed as
\begin{equation}
    \Pi_{\mu\nu}(k^2) \equiv g_{\mu\nu}\Pi_1(k^2)+ k_\mu\,k_\nu \Pi_2(k^2)
\end{equation}
where the scalar function $\Pi_1(k^2)$ characterizes the transverse component and contributes to the $S$-matrix, whereas the longitudinal component given by $\Pi_2(k^2)$ does not contribute when coupled to a conserved current. The mass counter-term for the massive dark photon is 
\begin{equation}
    \delta\, m_{A'}^2\,=\,\text{Re}[\Pi_1(m_{A'}^2)]
\end{equation}
and the wave function renormalization counter-term is
\begin{equation}
    \delta Z_{A'}\,=\,Z_{A'}-1 \simeq - \text{Re} \left[\frac{d\, \Pi_1 (m_{A'}^2)}{d\,k^2} \right]
\end{equation}
where $Z_{A'}$ is the wave function renormalization for a massive dark photon. The renormalized correction to the scalar part of the dark photon propagator is thus \cite{Jegerlehner:1990uiq}
\begin{equation}
    \Pi_{\rm ren}(k^2)\,=\,\Pi_1(k^2)\,-\,\delta\,m_{A'}^2\,+\,(k^2\,-\,m_{A'}^2)\,\delta Z_{A'}
\end{equation}
We discuss the one-loop contributions from the dark matter and the dark Higgs to the dark photon self-energy separately, below.  At NLO, these two classes of contributions are simply summed together.

\subsection{Contribution from Dark Matter}

Using dimensional regularization, the one-loop correction to $\Pi_1(k^2)$ with the DM fermion in the loop takes the form:
\begin{equation}
\begin{aligned}
     \Pi_1(k^2) &=  \frac{\alpha_D} {2 \pi \, (\epsilon -3)}\Bigg(\left(4\, m_\chi^2-(\epsilon -2)\, k^2\right) B_0\left(k^2;m_\chi^2,m_\chi^2\right) \\
    & +2 \,(\epsilon -2) A_0\left(m_\chi^2\right)\Bigg)
\end{aligned}
\label{eqn:Pi1}
\end{equation} 
where $\epsilon=\,4-D$. 

The dark matter contribution to the mass counter-term is 
\begin{eqnarray}
\label{eqn:dark-ph-mass-ren}
    \delta\, m_{A'}^2
    &=& \frac{\alpha_D} {2\pi \, (\epsilon -3)}\,\text{Re}\Bigg(
    2 (\epsilon -2) A_0\left(m_\chi^2\right) \\
    & & + \left(4\,m_\chi^2-m_{A'}^2 (\epsilon -2)\right) B_0\left(m_{A'}^2; m_\chi^2,m_\chi^2\right) \Bigg) \nonumber
\end{eqnarray}
and the DM contribution to the wave function renormalization counter-term $\delta Z_{A'}$ is
\begin{equation}
\begin{aligned}
    \delta\,Z_{A'}
    &= -\frac{\alpha_D}{2\pi (\epsilon -3)}  \,\text{Re}\Bigg((\epsilon -2) B_0\left(m_{A'}^2;m_\chi^2,m_\chi^2\right)\\
    & +\left(m_{A'}^2 (\epsilon -2)-4 m_\chi^2\right) B_0'\left(m_{A'}^2;m_\chi^2,m_\chi^2\right)\Bigg)
\end{aligned}
\end{equation}

\subsection{Dark Higgs Contribution}

The one-loop correction $\Pi_1(k^2)$ to the dark photon self-energy due to dark Higgs is
\begin{equation}
\begin{aligned}
    \Pi_1(k^2)&=\frac{\alpha_D\, q_\phi^2} {4\, \pi \, (\epsilon -3)\, k^2}\Bigg\{\Bigg(-2 k^2 \left(m_A^2 (2 \epsilon -5)+m_H^2\right)\\
    &+k^4+\left(m_A^2-m_H^2\right)^2\Bigg) B_0\left(k^2,m_A^2,m_H^2\right)\\
    &+A_0\left(m_H^2\right) \left((5-2 \epsilon ) k^2+m_A^2-m_H^2\right)\\
    &-A_0\left(m_A^2\right) \left(k^2+m_A^2-m_H^2\right)\Bigg\}
\end{aligned}
\end{equation}
where $\epsilon=\,4-D$. The contribution to the mass counter-term is
\begin{equation}
\begin{aligned}
    \delta m_{A'}^2&=\frac{\alpha_D\, q_\phi^2} {4\, \pi\,  m_{A'}^2 (\epsilon -3)}\text{Re}\Bigg\{\Bigg(-4 m_{A'}^4 (\epsilon -3)\\
    &-4 m_{A'}^2 m_H^2+m_H^4\Bigg) B_0\left(m_{A'}^2; m_{A'}^2,m_H^2\right)\\
    &-A\left(m_H^2\right) \left(2 m_{A'}^2 (\epsilon -3)+m_H^2\right)\\
    &+\left(m_H^2-2 m_{A'}^2\right) A_0\left(m_{A'}^2\right)\Bigg\}
\end{aligned}
\end{equation}
and the contribution to the wave function renormalization counter-term is

\begin{equation}
    \begin{aligned}
        \delta\,Z_{A'}\,&=\frac{\alpha_D\,q_\phi^2}{4\, \pi\,  m_{A'}^4 (\epsilon -3)} \text{Re}\Bigg\{m_H^2 \left(m_H^2-2 m_{A'}^2\right)\\ &B_0\left(m_{A'}^2;\,m_{A'}^2,m_H^2\right)\\
        &+m_{A'}^2 \left(4 m_{A'}^4 (\epsilon -3)+4 m_{A'}^2 m_H^2-m_H^4\right) B_0^\prime\left(m_{A'}^2;\,m_{A'}^2,m_H^2\right)\\
        &+\left(m_H^2-m_{A'}^2\right) A_0\left(m_{A'}^2\right)\\
        &+(m_{A'}-m_H) (m_{A'}+m_H) A_0\left(m_H^2\right)\Bigg\}
    \end{aligned}
\end{equation}

\bibliographystyle{apsrev4-1}
\bibliography{DM_loops}
\end{document}